\documentclass[doublecol]{epl2} 

\pdfoutput=1
\renewcommand{\revision}[1]{{#1}}

\newcommand{\bra}{\left\langle}
\newcommand{\ket}{\right\rangle}

\newcommand{\del}{\delta}

\def\pointx{\protect \includegraphics[height=1.6ex]{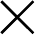}}%
\def\pointz{\protect \includegraphics[height=1.6ex]{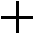}}%
\def\pointa{\protect \includegraphics[height=1.6ex]{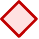}}%
\def\pointb{\protect \includegraphics[height=1.6ex]{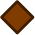}}%
\def\pointc{\protect \includegraphics[height=1.6ex]{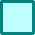}}%
\def\pointd{\protect \includegraphics[height=1.6ex]{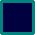}}%

\title{Instabilities and turbulence-like dynamics in an oppositely driven binary particle mixture}
\shorttitle{Lane formation, instabilities, and turbulence-like dynamics}

\author{\revision{Masahiro} Ikeda\inst{1} \and%
        \revision{Hirofumi} Wada\inst{2} \and%
        \revision{Hisao} Hayakawa\inst{1}}
\shortauthor{M. Ikeda \etal}

\institute{
\inst{1} Yukawa Institute for Theoretical Physics, Kyoto University, 606-8502 Kyoto, Japan\\
\inst{2} Department of Physics, Ritsumeikan University, Kusatsu 525-8577 Shiga, Japan
}


\abstract{
Using extensive particle-based simulations, we investigate out-of-equilibrium pattern dynamics in an oppositely driven binary particle system in two dimensions.
A surprisingly rich dynamical behavior including lane formation, jamming, oscillation and turbulence-like dynamics is found.
The ratio of two friction coefficients is a key parameter governing the stability of lane formation.
When the friction coefficient transverse to the external force direction is sufficiently small compared to the longitudinal one, the lane structure becomes unstable to shear-induced disturbances, and the system eventually exhibits a dynamical transition into a novel turbulence-like phase characterized by random convective flows.
We numerically construct an out-of-equilibrium phase diagram.
Statistical analysis of complex spatio-temporal dynamics of the fully nonlinear turbulence-like phase suggests its apparent reminiscence to the swarming dynamics in certain active matter systems.
}

\begin{document}

\maketitle

\section{Introduction}
Lane formation\cite{ivlev2012complex} is one of the representative examples of nonequilibrium phase transitions. 
When two kinds of particles are driven in opposite directions, the system exhibits a self-organization from a uniformly mixed state into strongly ordered anisotropic patterns. 
This driven segregation phenomenon, first found in the computer simulations\cite{PhysRevE.65.021402,PhysRevE.70.012401,RSC.FD.B202892C,Chakrabarti-EPL-2003,Liu2008224,epl-63-4-616,Delhommelle-PRE-2005}, has been observed in laboratory experiments such as mixtures of oppositely charged colloids\cite{nature03946,Vissers-SoftMatter-2011} or dusty plasmas in presence of an external electric field\cite{Sutterlin-PRL-2009}.
Another important class of examples is pedestrian and traffic flow dynamics\cite{RevModPhys.73.1067,PhysRevE.51.4282,PhysRevE.71.036121,Helbing-Nature-2000}.
By tracing single-particle motions in colloidal dispersions, a recent experiment has proposed the underlying mechanism of the lane formation as a dynamical ``lock-in'' state\cite{Vissers-SoftMatter-2011}.
A lateral mobility of particles is initially enhanced by frequent collisions, which however decreases considerably once lane is formed. 
This microscopic dynamics leads to the trapping of particles within the lanes and thus growth of the lane structures. 
This is also consistent with a physical picture employed in the phenomenological dynamic density-functional theory\cite{Chakrabarti-EPL-2003}.

\begin{figure}
\onefigure{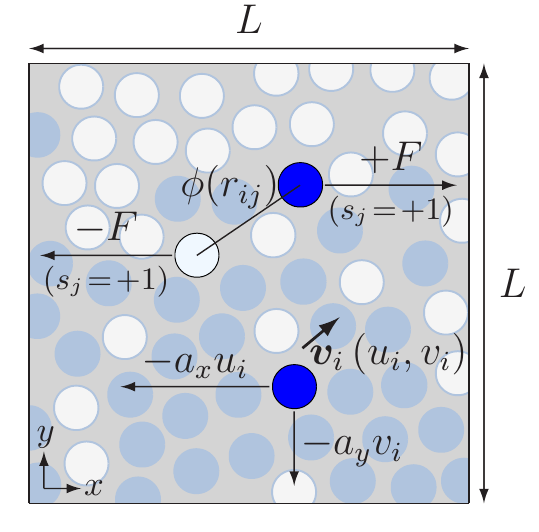}%
\caption{(Color online)    
An illustrative figure of our model. Blue particles are driven to the positive $x$ direction while white particles are driven to the negative $x$ direction. Each particle experiences a friction force proportional to its velocity, whose coefficients can be different in $x$ and $y$ directions.}
\label{fig:top}
\end{figure}

In this letter, using the particle-based simulations, we reveal entirely new aspects of such a lane forming system. 
We show that the anisotropy in the friction force is one of the key parameters that control the onset and stability of the lane formation. 
By systematically changing the ratio of two friction coefficients, we find surprisingly rich dynamics.
More specifically, lanes can become unstable for a sufficiently small transverse/longitudinal friction ratio. 
This transition appears to be triggered by long-wavelength interface undulations,
which eventually leads to turbulence-like convective flows. 
We numerically construct a stability diagram as functions of driving speed and the friction ratio, and identify the boundary lines separating several qualitatively different dynamical phases. 
In contrast to the lane phase, the structure of turbulence-like phase looks considerably isotropic due to random convective motions of clusters of the same particle species. 
However, statistical analysis reveals that the spatio-temporal order is characterized by correlation lengths that significantly exceed the constitutive particle size. 
Interestingly, these features are reminiscent of swarming of self-propelling elements such as in bacteria colonies\cite{Ramaswamy-AnnuRevCMP-2010,PhysRevLett.100.178103,Dombrowski-PRL-2004,Zhang-EPL-2009} and flocks of birds or fish\cite{Vicsek-PRL-1995,Shimoyama-PRL-1996}.

\section{Model}
An illustrative figure of the simulation setup is shown in fig.~\ref{fig:top}.
Our model system consists of $N$ particles interacting with a pair potential $\phi(r)$ (explained below) on a two-dimensional square plane of length $L$.
The particle number density is thus $\rho = N/L^2$. 
The system is driven to out-of-equilibrium state by an externally applied field $F$.
To describe how the particles react to $F$, we assign the variable $s$ for each particle such that the half of the particles with $s=+1$ are driven to the positive $x$ direction by $F$, while the other half with $s=-1$ are driven to the negative $x$ direction (see fig.~\ref{fig:top}).
Note that $s$ here is introduced as the label only to distinguish the type of responses of the particles to $F$.
In contrast to the Brownian dynamics studies~\cite{Chakrabarti-EPL-2003,PhysRevE.65.021402,PhysRevE.70.012401,RSC.FD.B202892C,epl-63-4-616}, we employ here a deterministic particle simulation with inertia, allowing the particles to possess arbitrary friction coefficients in the two directions, $a_x$ and $a_y$, respectively.
The friction terms are introduced here to mimic the momentum exchange with the substrate.
The equation of motion for the $i$-th particle of its position ${\bm x}_i(t)$ for $i=1,2,\cdots, N$ is
\begin{equation}
m\frac{d \bm v_i}{dt}=
-a_x (u_i-s_i V_0) \bm e_x
-a_y v_i \bm e_y
-\sum_{j \neq i} \frac{\partial \phi(|{\bm x}_i-{\bm x}_j|)}{\partial \bm x_i},
\label{eq:cov}
\end{equation}
where $d{\bm x}_i/dt={\bm v}_i=(u_i, v_i)$ is the particle velocity, $m$ is the particle mass common to all particles.
Without any interactions, the external force $F$ and the frictional force balance in steady state, which drives the particles toward their preferential directions at the constant speed $V_0=F/a_x$. 
Hereafter, we use $V_0$ to measure the strength of the external driving. 
To model the short-ranged repulsion between the particles {\it irrespective of} $s_i$, we employ the Weeks-Chandler-Andersen (WCA) potential\cite{Weeks-JChemPhys-1971}:
\begin{equation}
\phi_{\rm WCA}(r_{ij}) = 
4 \epsilon
\left \{
  \left (\frac \sigma {r_{ij}} \right )^{12}
- \left (\frac \sigma {r_{ij}} \right )^{6}
+ \frac 1 4
\right \},
\label{eq:wca}
\end{equation}
for $ r_{ij}\leq 2^{1/6} \sigma$ and otherwise $\phi_{\rm WCA}=0$, where $r_{ij}=|{\bm x}_i-{\bm x}_j|$.
We also tested the Lennard-Jones potential to evaluate effects of short-ranged attractions, which we found are negligible (data not shown).

Our model given in eq.~(\ref{eq:cov}) is largely motivated by pedestrian and traffic dynamics in which the self-propelling elements interact frictionally with the substrate, rather than thermal systems such as colloidal dispersions. 
The unique feature of the present model is the generalization of the friction coefficients $a_x$ and $a_y$. 
There are a number of examples on the anisotropic frictions such as elongated molecules moving in a viscous fluid~\cite{Happel-Brenner}.
In the context of pedestrian dynamics, $a_x$ and $a_y$ are often called ``sensitivity'' to the difference between its actual velocity and its optimal velocity $V_0$.
Pedestrians may try to keep their optimal velocity while they actively change their lanes, which implies that $a_y$ can be much smaller than $a_x$.   

To minimize effects of the boundaries, we impose the periodic boundary conditions for both $x$ and $y$ directions. 
We rescale all variables in the units of energy $\epsilon$, the particle diameter $\sigma$, and the particle mass $m$, which leads to the rescaled time $\tilde{t}=t \sqrt{m\sigma^2/\epsilon}$ and the rescaled velocity $\tilde{V}_0=V_0/\sqrt{\epsilon/m}$.
We discretize Eq.~(\ref{eq:cov}) with the time step $\Delta t$, and integrate it by using the velocity Verlet method.
To ensure a sufficient numerical accuracy, we choose $\Delta \tilde{t}=0.001$.
The initial particle configurations are prepared as follows.
Particles are first located randomly on the square lattice points.
The magnitude of the initial velocity assigned to each particle is fixed to be $|{\bm v}_i|=6\sqrt{\epsilon/m}$ with random orientations.
The initial configuration is then obtained by performing 1000 steps numerical integrations of Eq.~(\ref{eq:cov}) without introducing any dissipation (i.e., $a_x=a_y=0)$ in the absence of the external driving ($V_0=0$).

\section{Results}
\begin{figure}
\onefigure[width=0.99\linewidth]{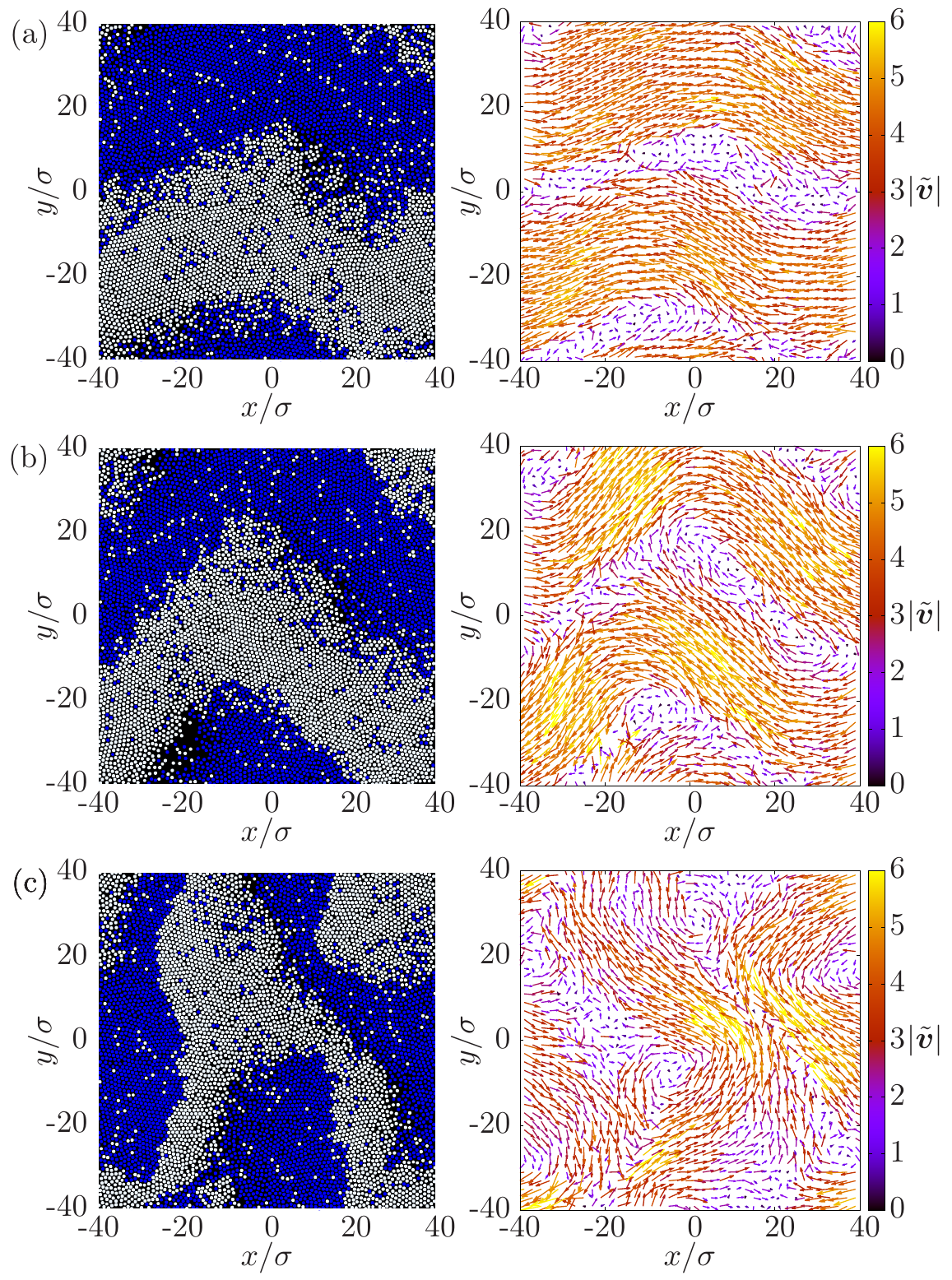}
\caption{(Color online)
Typical time evolution of a transiently formed lane structure in its unstable regime observed in our simulations. The transition to a turbulence-like flow proceeds from (a) to (c).
The parameters used in (a)-(c) are $\tilde{V}_0=5.0$, $a_y/a_x=0.2$, $L=80\sigma$ and $\rho\sigma^2=1.0$.
Left panel: snapshots of the particle configurations.
Right panel: the corresponding velocity field. Arrows show the orientation of the velocity and colors indicate the magnitude of the velocity. 
}
\label{fig:size_figs}
\end{figure}

For the isotropic friction case $a_y/a_x=1$, the uniformly mixed initial state becomes unstable and the usual lane formation proceeds. 
In contrast, for small transverse friction case $a_y/a_x \ll 1$, this lane formation is only transient and is unstable against long-wavelength undulations of the interfaces.
In fig.~\ref{fig:size_figs}, typical snapshots and the corresponding velocity field obtained from the simulations for $a_y/a_x=0.2$ are shown.
A transiently formed lane (fig.~\ref{fig:size_figs} (a)) becomes unstable and buckle (fig.~\ref{fig:size_figs} (b)), leading to the turbulence-like convective flow structure (fig.~\ref{fig:size_figs} (c)). 
The velocity fields shown on the right hand panel of fig.~\ref{fig:size_figs} (c) appear to be very complicated, where the complex transient jet or vortex structures are observed.
Statistical properties of this turbulence-like phase will be further investigated later on.
To our knowledge, this anisotropy-driven instability was previously unknown, and is the first main result of our present study.
While we performed the simulations for increasing system sizes, $L/\sigma=40, 80$ and 120 \revision{with the fixed particle number density ($\rho \sigma^2 = 1.0$)}, no qualitative and systematic changes were observed. 
We will therefore show the results obtained from the simulations for $L=80\sigma$ and $N=6400$ (i.e., $\rho\sigma^2=1.0$) below.

To characterize the observed patterns more quantitatively, we introduce the (local) temperature $T_A$ as
\begin{equation}
T_A = \overline{\bra
        \frac{m}{2}
        \left( {\bm v}_i - \overline{\bra {\bm v}_i \ket}_A \right)^2
      \ket}_A,
\label{eq:T-def}
\end{equation}
where $\overline O$ is the time average of $O$
between $\tilde{t}=5000$ and $10000$ for any variable $O$.
$A$ denotes a horizontal stripe area of $\sigma \times L$ in the simulation box
and $\bra O_i \ket_A = 1/N_A \sum_{i \in A} O_i$ for any variable $O_i$.
Here $N_A$ is the number of particles (white+blue) inside $A$.

\begin{figure}
\onefigure[width=0.99\linewidth]{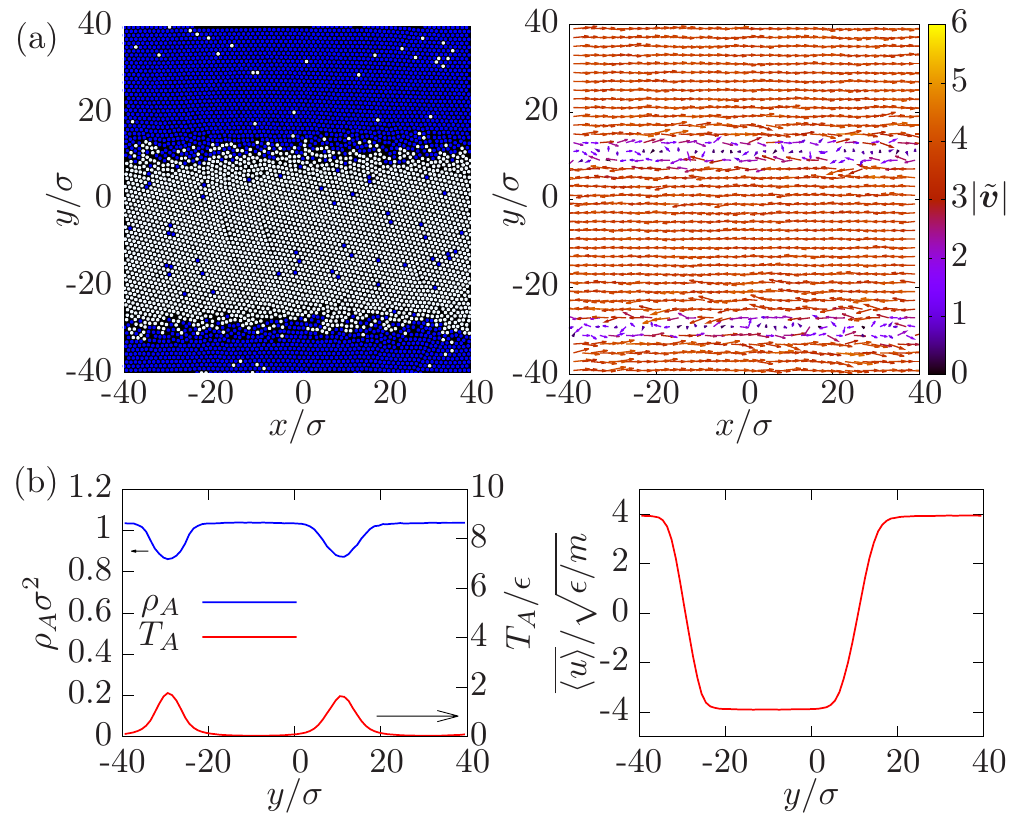}%
\caption{(Color online)
A typical snapshot (left panel of (a)) and the corresponding velocity field  (right panel of (a)) of lane configuration observed in our simulations.
\revision{Profiles of the total (white+blue) particle number density $\rho_A$ (left panel of (b)), the temperature $T_A$ (also left panel of (b)) and the $x$ component of the mean velocity (right panel of (b)) in the stable lane forming phase as functions of $y$.}
Simulation parameters used are $a_y/a_x=1.0, \tilde{V}_0=4.0, L=80\sigma$ and $\rho\sigma^2=1.0$.
}
\label{fig:tx_fig_a}
\end{figure}

Note that $T_A$ measures the kinetic energy in the local co-moving frame with its local mean velocity $\overline{\bra {\bm v} \ket}$, and is higher in the regions where collisions are more frequent such as in the interfaces between oppositely moving lanes.    
In fig.~\ref{fig:tx_fig_a}~(b), we show the spatial profile along $y$ direction of the local particle number density $\rho_A = \overline N_A / (\sigma L)$, the temperature $T_A$ and the $x$ component of the mean velocity $\overline{\bra u \ket}$ \revision{as a function of $y$ }in the stable laning phase obtained for $a_y/a_x=1.0$, $\tilde{V}_0=4.0$, $L=80\sigma$ and $\rho\sigma^2=1.0$. 

The temperature $T_A$ of the lane phase (shown in fig.~\ref{fig:tx_fig_a} (b)) is almost zero inside the lanes, whereas it is higher in the boundary layers between the oppositely oriented lanes.  
In contrast, the number density $\rho_A$ is lower in the boundary layers and is higher inside the lanes.
This is because there are much more collisions between the particles and considerably disordered particle configurations are realized within the layers.
For a sufficiently small $a_y/a_x$, the collision-induced shear stress at the boundary layer would be enough to induce collective migrations of the particles transverse to the flow direction, which ultimately causes the buckling\cite{Cubaud-PRL-2007,LeGrand-Piteira-PRL-2006} of the lanes (see fig~\ref{fig:size_figs}).

\section{Phase diagram}
In addition to the lane and turbulence-like phases, we also find a few other characteristic phases depending on $\tilde{V}_0$ and $a_y/a_x$, such as the completely jammed state in which clusters of particles push against each other without any motion nor transport of particles. 
To quantify the various patterns observed in our systems, we introduce the order parameter $\Phi$ as follows. 
First, we slice our simulation box into thin sections parallel to $x$ axis of width $\sigma$. Let $n_i^+$ and $n_i^-$ be the number of particles with $s=+1$ and $s=-1$, respectively, found in the $i$-th section, then we define our order parameter
\begin{eqnarray}
\Phi = \left\langle \frac{1}{N_s}\sum_{i=1}^{N_s}\frac{\vert n_i^+ - n_i^- \vert}{n_i^+ + n_i^-}\right\rangle,
\label{eq:def-phi}
\end{eqnarray}
where $N_s$ is the total number of the thin sections, and the bracket represents the long time average in steady state.
By definition, $\Phi$ changes from 0 to 1. For randomly mixed configurations, $\Phi$ is close to zero, while it approaches one as the system evolves the laning order.
We note that our definition of $\Phi$ is essentially the same as those introduced previously\cite{PhysRevE.65.021402}, and small differences are insignificant for describing the qualitative aspects of our results.

\begin{figure}
\onefigure[width=0.99\linewidth]{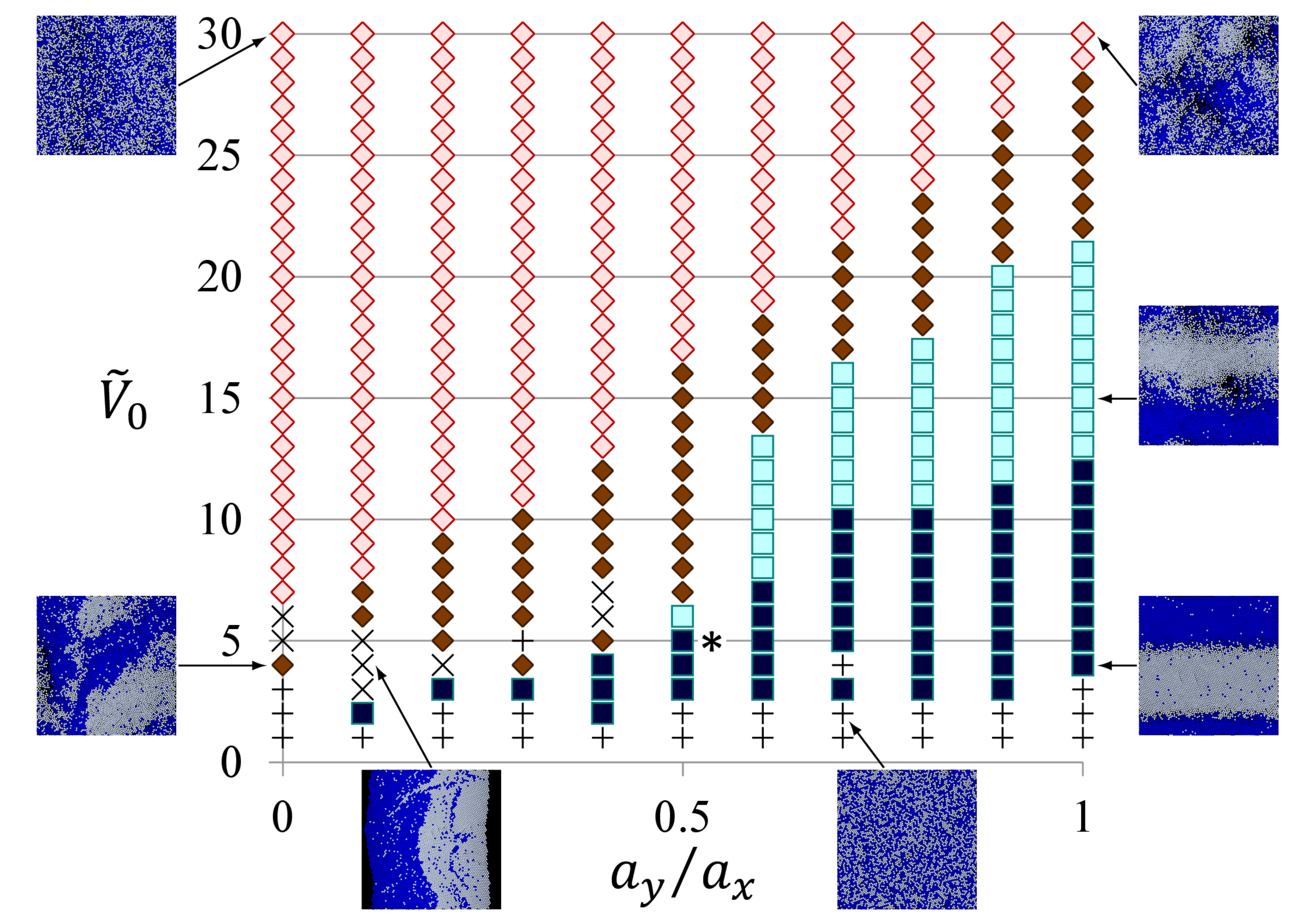}
\caption{(Color online)
The out-of-equilibrium phase diagram that is numerically constructed by looking at the order parameter $\Phi$ defined in eq.~(\ref{eq:def-phi}), on the plane spanned by the external driving speed $\tilde{V}_0$ and the friction constant ratio $a_y/a_x$.  The number density and the system size are fixed as $\rho\sigma^2=1.0$ and $L=80\sigma$.
The symbols on the diagram indicate the different values of $\Phi$:
\pointa : $\Phi$ is from 0.00 to 0.25;
\pointb : $\Phi$ is from 0.25 to 0.50;
\pointc : $\Phi$ is from 0.50 to 0.75;
\pointd : $\Phi$ is from 0.75 to 1.00.
Symbol \pointz\ and \pointx\ indicate non-flow phases:
\pointz : randomly mixed initial configuration;
\pointx : completely jammed state.
\revision{Note that the value of $\Phi$ in the non-flow phase is around 0.1.}
}
\label{fig:order}
\end{figure}

With the order parameter $\Phi$ defined above, we now construct a phase diagram. 
In fig.~\ref{fig:order}, the values of $\Phi$ obtained from our simulations are shown on the $\tilde{V}_0-(a_y/a_x)$ plane.
Five qualitatively different phases, the randomly mixed, stable lane, jammed, uniform gas, and convective or turbulence-like phases, are identified for a fixed number density $\rho\sigma^2=1.0$.
When $\tilde{V}_0$ is smaller than 2.0, the external force is insufficient to produce any macroscale particle flow, where the system almost keeps its initial configuration.
Once $\tilde{V}_0$ becomes larger than 2.0, particles start to move and the large-scale pattern formation proceeds.  
For $V_0$ slightly larger than its critical value 2.0, the usual lane forming dynamics is observed if the dissipation is almost isotropic, for which we obtain $\Phi \approx 1$. 
As $V_0$ further increases, the kinetic temperature $T$ increases significantly, leading to a uniform gas-like phase. 
This process is characterized by a gradual broadening of the lane interfaces, and we note that this gas-like phase was addressed previously in Ref.~\cite{Liu2008224}. 
However, the lane destabilization observed in our system is distinct from this gas phase, because it is induced only by the strong anisotropy in the friction coefficients.
Also, the resulting turbulence-like phase is characterized by the collective movements of the particles, rather than the gas-like mixture. 
\revision{As $V_0$ increases, $\Phi$ decreases monotonically for $a_y/a_x \approx 1$, but for smaller $a_y/a_x$ (typically less than 0.5), $\Phi$ changes almost discontinuously from the lane to the turbulence-like, from the lane to the jammed, and from the jammed to the turbulence-like phases (see Fig.4).
While the former is the continuous crossover, the latter may arise from a certain hydrodynamic-type instability.}

At several points indicated by ``\pointx'' in fig.~\ref{fig:order}, $\Phi$ is exceptionally small as compared with those of the neighboring points.
These points correspond to the jammed phase, where the particles do segregate but their interface appears {\it perpendicular to} the driving direction. In this case, the whole system maintains its static force balance state without any particle currents.
As $V_0$ is increased, the vertical interface becomes unstable against long wavelength undulations, and the particles start to move.

\begin{figure}
\onefigure{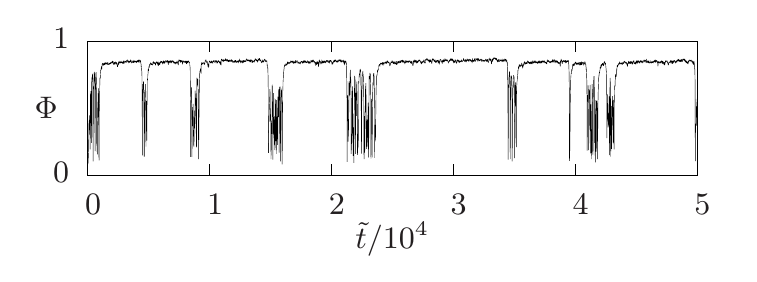}
\caption{(Color online)
Time evolution of the order parameter $\Phi$ during the extensive long time simulation for $\tilde{V}_0=5$ and $a_y/a_x=0.5$. In this case, intermittent lane formations are observed between laning phase and the turbulence-like convective flow phase.}
\label{fig:periodic}
\end{figure}

{\it Turbulence-like phase--}
As $a_y/a_x$ is decreased, the system begins to show complex spatio-temporal dynamics.
For example, at a point indicated by * (as well as its small surrounding region) on the diagram in fig.~\ref{fig:order}, we observe a unique ``quasi-periodic'' time evolution of $\Phi$, which is shown in fig.~\ref{fig:periodic}.
The lane structure (with $\Phi\approx 1$)  breaks up stochastically, but the resulting disorganized state ($\Phi\approx 0$) is unstable and the lane formation immediately proceeds again, resulting in the spikes of $\Phi$ in fig.~\ref{fig:periodic}.
Strikingly, although the precise mechanism of this nonlinear oscillation is at present unclear, this whole process was observed to repeat during our extensive long time simulations, and not transient.

\begin{figure}
\onefigure{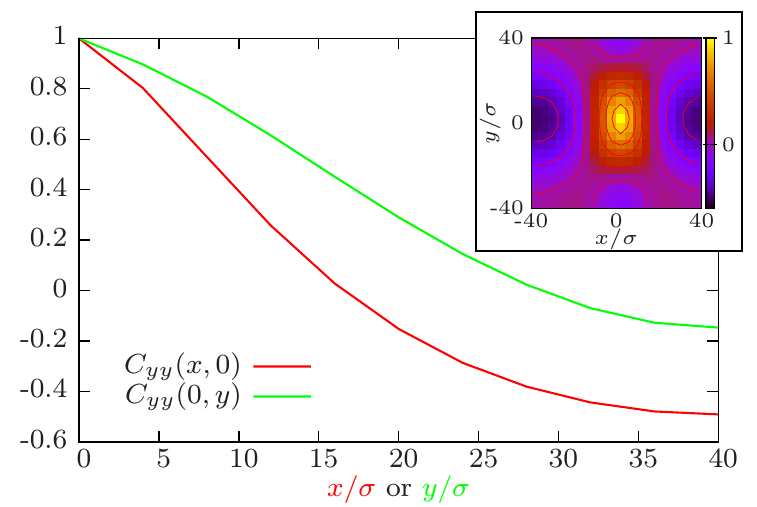}%
\caption{(Color online)
Spatial correlation function of the $y$-component of the instantaneous velocity field obtained from the simulations for $a_y/a_x=0.0$ and  $\tilde{V}_0=4.0$.
In the main panel, $C_{yy}(x,y=0)$ (red and solid line) and $C_{yy}(x=0,y)$ (green and dashed line) are plotted. Inset figure shows the two-dimensional density plot of $C_{yy}$ to highlight the anisotropic feature of the velocity correlations.}
\label{fig:correlation}
\end{figure}

For even smaller values of $a_y/a_x$, the system is brought into a highly chaotic or turbulence-like flow phase, which contrasts to the highly ordered and anisotropic lane phase.
To quantitatively characterize this phase, we look at the correlation function of $y$-component instantaneous velocity field, which is defined as
\begin{equation}
 C_{yy}(x, y)
 = \frac{\langle v(x_0,y_0,t) v(x_0+x,y_0+y,t) \rangle}
        {\langle v(x_0,y_0,t) v(x_0,y_0,t) \rangle},
 \label{eq:correlation}
\end{equation}
where $\langle \cdots \rangle$ represents average over time $t$ and space $x_0,y_0$.
Whereas the external force is absent in $y$ direction, the correlation function is strongly anisotropic, as seen from the inset of fig.~\ref{fig:correlation}. 
This feature is also understood by comparing the profiles of $C_{yy}$ along the $x$-axis (with $y=0$) and along the $y$-axis (with $x=0$), shown in the main panel of fig.~\ref{fig:correlation}. 
The transverse correlation $C_{yy}(x,y=0)$ becomes negative at around $x_c/\sigma \approx 16$, while the longitudinal one $C_{yy}(x=0,y)$ decays exponentially from the origin and becomes slightly negative at around $y_c/\sigma\approx 30$. 
This observation provides us a physical picture that densely packed particles move in a coordinated manner and form an elongated cluster or jet of typical sizes $x_c$ and $y_c$.
In addition, the clusters of the oppositely moving directions encounter in several places, as the upward and downward flow are neighboring in fig.~\ref{fig:size_figs}, which gives rise to transient vortex structures.
While external driving is solely along the $x$ direction, the particles find their escape ways to the less-dissipating $y$-direction. 
Thus the energy is transmitted via the anisotropic dissipation from the $x$ to $y$ direction, which gives rise to the distinct collective dynamics at intermediate length scales (see fig.~\ref{fig:correlation}).
A rich variety of the extended spatio-temporal dynamics in our lane forming system is the second main result of this paper.

\section{Discussion}
Here we outline simple energetic arguments for the stability of lane structures, which might provide some (but rough) physical interpretations on the observed instabilities.
The total energy of our system consists of the kinetic and potential energies and is written as
\begin{eqnarray}
  E = \sum_{i=1}^N
  \left[
      \frac{m {\bm v}_i^2}{2}+\sum_{j > i} \phi(r_{ij})
  \right]
\end{eqnarray}
Now we introduce the drift velocity $u_{\rm d} = 1/N \sum_{i=1}^{N} s_i u_i$ and fluctuations from it as $\delta u_i = u_i - s_i u_{\rm d}$. 
Note that $\delta v_i = v_i$ because we can obtain from eq.~(\ref{eq:cov}) that $\sum_{i=1}^{N} v_i=0$ in steady state.
With these quantities, we can compute the time derivative of the total energy $\dot{E}=dE/dt$ as
\begin{equation}
   \dot{E}
=
 \revision{ N a_x u_{\rm d} ( V_0 - u_{\rm d} ) }
 - \sum_{i=1}^N \left[ a_x(\delta u_i)^2+a_y(\delta v_i)^2\right].
\label{eq:ene}
\end{equation}
\revision{The first term on the right hand side of eq.~(\ref{eq:ene}), $\dot{W}=N a_x V_0 u_{\rm d}$, represents the work done by the external force per unit time. The second term, $\dot{Q}_{\rm hk}=N a_x u_{\rm d}^2$, known as the rate of housekeeping heat\cite{PTPS.130.29}, corresponds to the average energy dissipation necessary to keep the whole system at this steady state.} 
Third terms in eq.~(\ref{eq:ene}),  $\dot{Q}_{\rm ex}=\sum_{i=1}^N \left[a_x(\delta u_i)^2+a_y (\delta v_i)^2\right]$, represents the ``excess'' energy dissipation per unit time, which arises from fluctuations of particle velocities from its steady-state average.  
Equation~(\ref{eq:ene}) suggests the usual energy balance (per unit time), $\dot{E}=\dot{W}-\dot{Q}$, where $Q=Q_{\rm hk}+Q_{\rm ex}$. 

Whereas our model does not have any direct route to thermal equilibrium, the particle motions are under-damped and the system is expected to be fully thermalized at a driven steady state owning to their inertia.
Given the principle of minimum entropy production in such driven steady state (or something analogous to this), the particle configurations that can minimize $Q_{\rm ex}$ should appear as the most stable one.
When both $a_x$ and $a_y$ have large enough positive values, it is achieved by minimizing $\del u_i$ and $\del u_i$, which indicates that the system prefers a stable lane formation.
However, when $a_y$ becomes sufficiently small, velocity fluctuations along $y$ direction can be unboundedly large, which could ultimately lead to a buckling of lane structures.
Whereas the validity of the minimum entropy production principle to our system is entirely unclear at present~\footnote{%
We note that there is no universal principle for the pattern selections in dissipative structures even when a system is close to (thermal) equilibrium. See for example ref.~\cite{Landauer-PRA-1975}.}, such a physical picture is fully consistent with what we have found in the simulations and could be useful to formulate a more elaborate theory in future studies.

Finally, we briefly remark on the turbulence-like phase. This dynamics is governed by the highly coordinated motions of the clusters of the same particle species.
The instantaneous velocity field exhibits complex vortex and jet structures, whose typical sizes considerably exceed the constitutive particle size (see fig.~\ref{fig:size_figs}).
These features present some similarities with those of swarming phase found in bacterial suspensions or colonies~\cite{Zhang-EPL-2009,Dombrowski-PRL-2004}. 
In active matter systems, however, there are uptake of free energy and the systematic movement under the ``force-free'' condition at the level of each active elements~\cite{Ramaswamy-AnnuRevCMP-2010}. 
Such a ``micro-to-macro'' energy transport is characteristic in active matter, and obviously contrasts to our lane forming system where the energy input is solely at the macroscopic scale.
Nevertheless, this apparent similarity might suggest a certain underlying mechanics that is common to driven soft matter and active matter systems.
Elucidating the microscopic particle dynamics in our system would provide more quantitative insights into the observed instability, which will also help develop the linear and nonlinear stability analyses.

\section{Summary}
Our study opens up the entirely new aspects of the lane forming system, i.e., the prototype model for non-equilibrium phase transitions, with a wealth of novel complex dynamical behavior.
The results obtained here will be potentially important not only in technological applications such as microfluidics, but also for understanding collective human behavior such as crowd controls or panic dynamics in densely populated areas~\cite{Helbing-Nature-2000}.

\acknowledgments
One of the authors (M. I.) thanks S. Yabunaka for discussions.
This work was supported by the Global COE Program
``The Next Generation of Physics, Spun from Universality \& Emergence''
from the Ministry of Education, Culture, Sports, Science and Technology
(MEXT) of Japan.
Numerical computation in this work was carried out
at the Yukawa Institute Computer Facility.

\bibliographystyle{eplbib}
\bibliography{lane.bib}

\end{document}